\documentclass[11pt]{article}
\usepackage{amssymb}
\usepackage{citesort}
\usepackage{url}
\usepackage{cite}
\usepackage{ntheorem}
\newcommand{\mcN}{{\mycal N}}

\newcommand{\hOmega}{\widehat{\Omega}}

\newcommand{\riemgz}{g_0}

\renewcommand{\hbar}{{\overline \riemgz}}

\newcommand{\mcD}{{\mycal D}}

\newcommand{\nablash}{\nabla{\kern -.75 em
     \raise 1.5 true pt\hbox{{\bf/}}}\kern +.1 em}
\newcommand{\Deltash}{\Delta{\kern -.69 em
     \raise .2 true pt\hbox{{\bf/}}}\kern +.1 em}
\newcommand{\Rslash}{R{\kern -.60 em
     \raise 1.5 true pt\hbox{{\bf/}}}\kern +.1 em}

\newcommand{\mcU}{{\mycal U}}

\newcommand{\hyp}
{{\mycal S}}

\newcommand{\mcM}{{\mycal M}}
\newcommand{\mcK}{{\mycal K}}

\newcommand{\bea}{\begin{eqnarray}}
\newcommand{\beaa}{\begin{eqnarray*}}
\newcommand{\bean}{\begin{eqnarray}\nonumber}

\newcommand{\bel}[1]{\begin{equation}\label{#1}}
\newcommand{\beal}[1]{\begin{eqnarray}\label{#1}}
\newcommand{\beadl}[1]{\begin{deqarr}\label{#1}}
\newcommand{\eeadl}[1]{\arrlabel{#1}\end{deqarr}}
\newcommand{\eeal}[1]{\label{#1}\end{eqnarray}}
\newcommand{\eead}[1]{\end{deqarr}}
\newcommand{\eea}{\end{eqnarray}}
\newcommand{\eeaa}{\end{eqnarray*}}

\newcommand{\Ricc}{\mathrm{Ric}\,}

\newcommand{\be}{\begin{equation}}
\newcommand{\ee}{\end{equation}}

\newcommand{\tr}{\mbox{\rm tr}}

\newcommand{\eq}[1]{\eqref{#1}}

\DeclareFontFamily{OT1}{rsfs}{}
\DeclareFontShape{OT1}{rsfs}{m}{n}{ <-7> rsfs5 <7-10> rsfs7 <10->
rsfs10}{} \DeclareMathAlphabet{\mycal}{OT1}{rsfs}{m}{n}

\usepackage{amsmath} 

\let\ssection=\section

\renewcommand{\section}{\setcounter{equation}{0}\ssection}

\newtheorem{defi}{\sc Definition\rm}[section]

\newtheorem{Theorem}[defi]{\sc Theorem\rm}

\newtheorem{corollary}[defi]{\sc Corollary\!\rm}
\theorembodyfont{\upshape}

\newcommand{\qed}{\hfill $\Box$\bigskip}

\def \Reel{\mathbb{R}}

\def \R {\Reel}

\def \Nat{\mathbb{N}}

\def \N {\Nat}

\newcommand{\Sp}[0]{\mbox{$\mathbb{S} $}}

\newcounter{mnotecount}[section]

\newcommand{\rmnote}[1]{}

\newcommand{\dive}{{\mbox{div\,}}}

\newcommand{\gt}{\gamma_{T}}
\newcommand{\Gt}{{\Gamma}_{T}}

\newcommand{\mut}{\mu_{T}}
\newcommand{\tmut}{{\tilde\mu}_{T}}

\newcommand{\pt}{\psi_{T}}
\newcommand{\tpt}{{\tilde \psi}_{T}}

\newcommand{\calN}{\mathcal N}
\newcommand{\calL}{\mathcal L}
\newcommand{\calD}{{\mathcal D}}
\newcommand{\X}{{\bf X}}
\newcommand{\Y}{{\bf Y}}

\newtheorem{definition}{Definition}

\begin{document}
\title{Initial data engineering}
\author{
Piotr T. Chru\'sciel\thanks{Partially supported by a Polish
Research Committee grant 2 P03B 073 24; email \protect\url{
piotr@gargan.math.univ-tours.fr}, URL {
www.phys.univ-tours.fr/\,$\tilde{}$\,piotr}}
\\
Universit\'e de Tours \and James Isenberg\thanks{Partially
supported by the NSF under Grants PHY-0099373 and PHY-0354659;  
email \protect\url{jim@newton.uoregon.edu}, URL:
{www.physics.uoregon.edu/\,$\tilde{}$\,jim}}
\\ University of Oregon
\and Daniel Pollack\thanks{Partially supported by the NSF under
Grant DMS-0305048 and the UW Royalty Research Fund;  email
\protect\url{ pollack@math.washington.edu}, URL:
www.math.washington.edu/\,$\tilde{}$\,pollack}
\\
University of Washington
}
\date{}

\maketitle

\begin{abstract}
We present a local gluing construction for
general relativistic initial data sets. The method applies to
\emph{generic} initial data, in a sense which is made precise. In
particular the trace of the extrinsic curvature is not assumed to
be constant near the gluing points, which was the case for
previous such constructions. No global conditions on the initial
data sets such as compactness, completeness,  or asymptotic
conditions are imposed. As an application, we prove
existence of spatially compact, maximal globally hyperbolic,
vacuum space-times without any closed  constant
mean curvature spacelike hypersurface.
\end{abstract}

\section{Introduction}\label{Sintro}

Let $(M_a,\gamma_a,K_a)$, $a=1,2$,  be two (arbitrary dimensional) 
general relativistic
initial data sets; by this we mean that each $\gamma_a$ is a
Riemannian metric on the n dimensional manifold $M_a$, while each $K_a$ is a
symmetric two-covariant tensor field on $M_a$. 
Such a data set is called vacuum data if it satisfies the vacuum
Einstein constraint equations
\begin{eqnarray}
\label{ce20n0vac}
R(\gamma) -(2\Lambda+ |K|_\gamma^2-(\tr_\gamma K)^2)= 0\\
D_i(K^{ij}-\tr_\gamma K \gamma^{ij})=0
\label{ce20n1vac}
\end{eqnarray}
 where $R(\gamma)$ is the scalar curvature (Ricci scalar) of the
 metric $\gamma$, and $\Lambda$ is the cosmological constant.
 The \emph{vacuum local gluing
problem} can be formulated as follows:

\begin{quote}
Let $p_a\in M_a$ be two points, and let $M$ be the manifold
obtained by removing from $M_a$ geodesic balls of radius
$\epsilon$ around the $p_a$, and gluing in a neck $\Sp^{n-1}\times I$,
where $I$ is an interval. Can one find  vacuum initial data
$(\gamma,K)$ on $M$ which  coincide with the original vacuum data
away from a small neighborhood of the neck?
\end{quote}

It is natural to pose the same question in field theoretical
models with matter. A formulation that avoids the issue of
specifying the precise nature of the matter fields is obtained if
we represent these fields in the initial data set by the matter
energy density function $\rho$ and the matter energy-momentum
vector $J$, requiring that they satisfy the dominant energy
condition\footnote{Recall that \eq{dec} might fail when quantum
phenomena are taken into account. We note that the local gluing
problem is trivial if no energy restrictions, or matter content
restrictions, are imposed, as then both the metric and the
extrinsic curvature can be glued together in many different ways.}
\bel{dec} \rho\ge |J|\;.\ee
The Einstein-matter constraints then relate $\rho$ and $J$ 
to the gravitational fields via the following
 \beal{ce20n} & {16 \pi \rho = R(\gamma) -(2\Lambda+
|K|_\gamma^2-(\tr_\gamma K)^2)}\;, &
 \\ & 16 \pi J^j = 2 D_i(K^{ij}-\tr_\gamma K \gamma^{ij})\;.
&\label{ce20n1}\eea

As a  variation of the local gluing problem, one has the 
\emph{wormhole creation problem},
or the \emph{vacuum wormhole creation} problem: one starts with a
single initial data set $(\tilde M, \tilde \gamma, \tilde K)$, and
chooses a pair of points $p_a \in \tilde M$. As before one forms the new
manifold $M$ by replacing small geodesic balls around these points
by a neck $S^{n-1}\times I$, and one
asks for the existence of initial data on $M$ which satisfy either  the
vacuum or the Einstein-matter constraints, and which coincide with the
original data away from the neck region.

It is easily seen that for certain special sets of initial data,
such constructions
are not possible: consider, for example, the flat initial data set
$(\R^3,\delta,0)$ associated with Minkowski space-time. It follows
from the positive energy theorem that this set of data cannot be glued
to any data on a compact manifold without globally perturbing
the metric (so that the mass is nonzero).

The object of this work is to show that the above gluing constructions 
can be performed for \emph{generic} initial data sets. To make our notion
of genericity precise, some terminology is needed.  Let $P$ denote
the linearisation of the map which takes a set of data  $(g,K)$ to
the constraint functions appearing in
\eq{ce20n0vac}-\eq{ce20n1vac}, and let $P^*$ be its formal
adjoint. Recall that a \emph{Killing Initial Data} (KID) is
defined as a solution $(N,Y)$ of the set of equations
$P^*(N,Y)=0$. These equations are  given explicitly by
\be
\label{4} 0=\left(
\begin{array}{l}
2(\nabla_{(i}Y_{j)}-\nabla^lY_l g_{ij}-K_{ij}N+\tr K\; N g_{ij})\\
 \\
\nabla^lY_l K_{ij}-2K^l{}_{(i}\nabla_{j)}Y_l+
K^q{}_l\nabla_qY^lg_{ij}-\Delta N g_{ij}+\nabla_i\nabla_j N\\
\; +(\nabla^{p}K_{lp}g_{ij}-\nabla_lK_{ij})Y^l-N \Ricc(g)_{ij}\\
\; +2NK^l{}_iK_{jl}-2N (\tr \;K) K_{ij}
\end{array}
\right)\;.\ee  We shall denote by $\mcK(\Omega)$ the set of KIDs
defined  on an open set $\Omega$ (note that we impose no boundary
conditions on $(N,Y)$.)
In a vacuum space-time $(\mcM,g)$ (possibly with non-zero
cosmological constant) the KIDs on a spacelike hypersurface $\Omega$
are in one-to-one correspondence with the Killing vectors of
$g$ on the domain of dependence of $\Omega$~\cite{Moncrief75}. A
similar statement, with an appropriately modified equation for the
KIDs, holds in electro-vacuum for appropriately invariant
initial data for the gravitational and electromagnetic fields. 
(The reader is referred
to~\cite{ChBeigKIDs} for comments about such data for general matter
fields.)

We note that the gluing problem is in fact  a special case of the wormhole
creation problem if one allows  $\tilde M$ to be a non connected manifold.
Hence from now on we shall  assume that $\tilde M$ has
either one or two components, with $p_a\in \tilde M$, $a=1,2$. The
first main result of our paper concerns vacuum initial data:

\begin{Theorem}
\label{Tlgluingv} Let $(\tilde M,\tilde \gamma,\tilde K)$ be a
smooth vacuum initial data set, and consider two open sets
$\Omega_a\subset \tilde M$ with compact closure and smooth
boundary such that $$\mbox{ the set of KIDs, $\mcK(\Omega_a)$, is
trivial.} $$ Then for all  $p_a\in \Omega_a$, $\epsilon >0$ and
$k\in \N$ there exists a smooth vacuum initial data set
$(M,\gamma(\epsilon),K(\epsilon))$ on M such that
$(\gamma(\epsilon),K(\epsilon))$ is $\epsilon$-close to  $(\tilde
\gamma,\tilde K)$ in a $C^k\times C^k$ topology away from
$B(p_1,\epsilon)\cup B(p_2,\epsilon)$. Moreover
$(\gamma(\epsilon),K(\epsilon))$ coincides with $(\tilde
\gamma,\tilde K)$ away from $\Omega_1\cup \Omega_2$.
\end{Theorem}

 The hypothesis of smoothness has been made for
simplicity. Similar results, with perhaps some finite loss in
differentiability, can be obtained for initial data sets with
finite H\"older or Sobolev differentiability.

Some comments
about the no-local-KIDs condition $\mcK(\Omega_a)=\{0\}$  are in
order. As noted above,  this is equivalent to the condition that there
are no  Killing vectors defined on the domain of dependence of the regions
$\Omega_a$ in the associated vacuum space-time. First, the
result is sharp in the following sense: as discussed above,
 initial data for Minkowski space-time cannot locally be glued
to anything which is non-singular and vacuum. This meshes with the fact that for  Minkowskian initial
data, we have $\mcK(\Omega)\ne\{0\}$ for any open set $\Omega$. Next, it is
intuitively clear that for generic space-times there will be no
locally defined Killing vectors, and several precise statements to
this effect have been proved in \cite{ChBeignokids}. Thus, our
result can be interpreted as the statement that for generic vacuum
initial data sets the local gluing can  be performed around
arbitrarily chosen points $p_a$. In particular it follows from the
results in \cite{ChBeignokids} that the collection of initial data
with generic regions $\Omega_a$ satisfying the hypotheses of
Theorem~\ref{Tlgluingv} is not empty.
Further, it follows from the results here together with those in~\cite{IMP}
and in~\cite{ChBeignokids} that the following initial data sets can always be
glued together, near arbitrary points, after a (perhaps global) perturbation
which is $\epsilon$-small away from the gluing region:
\begin{itemize}
\item initial data containing an asymptotically flat region
\item initial data containing a conformally compactifiable CMC region
\item CMC initial data on a compact boundaryless manifold.
\end{itemize}

Let us denote by $\mcN(\Omega)$ the set of functions $N$
satisfying, on $\Omega$, the second of equations \eq{4} with
$K\equiv 0$. Theorem~\ref{Tlgluingv} has the following purely
Riemannian ``time-symmetric" counterpart:

\begin{Theorem}
\label{Tlgluingvr} Let $(\tilde M,\tilde \gamma)$ be a smooth
Riemannian manifold with non-positive constant scalar curvature
$\nu$, and consider two open sets $\Omega_a\subset \tilde M$ with
compact closure and smooth boundary such that
$$\mcN(\Omega_a)=\{0\}\;. $$ Then for
all  $p_a\in \Omega_a$, $\epsilon >0$ and $k\in \N$ there exists a
Riemannian manifold $(M,\gamma(\epsilon))$ with scalar curvature
$\nu$ such that $\gamma(\epsilon)$ is $\epsilon$-close to $\tilde
\gamma$ in a $C^k$ topology away from $B(p_1,\epsilon)\cup
B(p_2,\epsilon)$. Moreover $\gamma(\epsilon)$ coincides with
$\tilde \gamma$ away from $\Omega_1\cup \Omega_2$.
\end{Theorem}

The proof is a simplified version of that of
Theorem~\ref{Tlgluingv}; we leave the details to the reader. This
result is the local counterpart of the gluing theorem of
Joyce~\cite{Joyce} for $\nu<0$, and appears to be 
completely new in the case $\nu=0$.

A noteworthy consequence of Theorem~\ref{Tlgluingv}, proved in
Section~\ref{Sex}, is the following:

\begin{corollary}
\label{Tex} There exist  vacuum maximal globally hyperbolic
space-times with compact Cauchy surfaces which contain no compact
boundaryless spacelike hypersurfaces with constant mean curvature.
\end{corollary}

 It is clear that there exist equivalents of
Theorem~\ref{Tlgluingv} in non-vacuum field theoretical models.
However,  proofs for such models require a case-by-case
analysis of the corresponding gluing and KID equations. It is
therefore noteworthy that one can make a general statement
assuming only the dominant energy condition, which we use in its
$(n+1)$-dimensional formulation:
\bel{dec4} T_{\mu\nu}X^\mu Y^\nu \ge0 \ \mbox{ for all timelike
future directed vectors $X^\mu$ and $ Y^\mu$}\;.\ee The second
main result of this paper reads:

\begin{Theorem} \label{Tlgluingnv}
Consider a smooth solution  $(\mcM, g)$ of the Einstein field
equations $G_{\mu\nu}=8\pi T_{\mu\nu}$, with  one or
two connected components, and with matter fields satisfying  the
dominant energy condition \eq{dec4}.  Let $\tilde M$ be a
spacelike hypersurface in $\mcM$ with induced data $(\tilde
\gamma,\tilde K)$,
and let $$\mbox{ $p_a\in \tilde M$, $a=1,2$, be two points at
which the inequality \eq{dec4} is strict.} $$ Then for all
$\epsilon
>0$ there exists a smooth initial data set
$(M,\gamma(\epsilon),K(\epsilon))$ on M satisfying the dominant
energy condition such that $(\gamma(\epsilon),K(\epsilon))$
coincides with $(\tilde \gamma,\tilde K)$ away from
$B(p_1,\epsilon)\cup B(p_2,\epsilon)$.
\end{Theorem}

 The reader will have observed that
Theorem~\ref{Tlgluingv} concerns initial data sets only, while in
Theorem~\ref{Tlgluingnv} the starting point is a space-time. This
is related to the fact that we have not made any assumptions on the
matter fields except energy dominance.
If $(\tilde M, \tilde g, \tilde K)$ has constant mean curvature,
then the proof of Theorem~\ref{Tlgluingnv} is such that we could
restate the result  purely in terms of initial data, with no reference
to the space-time  $(\mcM, g)$.

Theorems~\ref{Tlgluingv} and \ref{Tlgluingnv} are established in
Section~\ref{Sgluing}. The proofs are a mixture of gluing
techniques developed in~\cite{IMaxP,IMP,IMP2} and those
of~\cite{CorvinoSchoenprep,Corvino,ChDelay}. In fact, the proof
proceeds via a generalisation of the analysis in~\cite{IMP,IMP2}
to compact manifolds with boundary; this is carried through in
Section~\ref{Sbound} in vacuum with cosmological constant
$\Lambda=0$, and in Section~\ref{Scsnv} with matter and
$\Lambda\in\R$.
These results may be of independent interest.  In order to 
have  CMC initial data near the gluing points, which the analysis
based on \cite{IMP} requires,  we make use of the work of
Bartnik \cite{bartnik:variational} on the plateau problem
for prescribed mean curvature spacelike hypersurfaces in a Lorentzian
manifold.

\bigskip

{\noindent \sc Acknowledgements:}  We  acknowledge
support from the Centre de Recherche Math\'ematiques, Universit\'e
de Montr\'eal, and the American Institute of Mathematics,
Palo Alto,
where the final stages of work on this paper were carried out.
PTC and JI are also grateful to the
Mathematics Department of the University of Washington and its members
for their friendly hospitality.

\section{The (global) gluing construction for the vacuum
constraints with $\Lambda=0$ for manifolds with boundaries}
\label{Sbound} 

In this section we formulate some generalisations of the results
in~\cite{IMP} and~~\cite{IMP2} to vacuum initial data sets on
manifolds with boundary, with vanishing cosmological constant; the
vacuum case with $\Lambda\ne 0$ will be covered by the analysis in
Section~\ref{Scsnv}. Although the paper \cite{IMP} only treats the
case $n=3$, since the generalization to higher dimensions is not
difficult (the necessary modifications are discussed in
\cite{IMaxP}),  we work here in general dimension $n\geq3$.
We begin with an initial data set $(\tilde M,
\tilde \gamma, \tilde K)$ where $\tilde M$ has non-empty smooth
boundary $\partial \tilde M$ and we assume first that $\tilde K$
has constant trace $\tilde \tau =\tr_{\tilde \gamma} \tilde K$
(i.e., these are constant mean curvature, or CMC, initial data
sets). Decomposing $\tilde K$ into its trace and trace-free
components, we write $\tilde K = \tilde \mu + \frac{\tilde
\tau}{n}\tilde \gamma$. Since $\tr_{\tilde \gamma}\tilde  K$ is
constant, the vacuum momentum constraint equation implies that
$\tilde \mu$ is divergence free as well (\emph{i.e.},\/ $\tilde
\mu$ is a transverse--traceless tensor). We ``mark'' $\tilde M$
with two points $p_a$, $a=1,2$,  about which we will perform the
gluing. The global gluing construction can be carried out in this setting,
with Dirichlet boundary conditions on the perturbation terms which
arise in applying the conformal method (see (\ref{vlapb}) and 
(\ref{lich-bvp}) below), generalizing the result of~\cite{IMP}.

\begin{Theorem}
\label{Tggluing} Let  $(\tilde M, \tilde \gamma, \tilde K; p_a)$
be a smooth, marked, constant mean curvature solution of the
Einstein vacuum
 constraint equations with cosmological constant $\Lambda=0$
on $\tilde M$, an $n$-manifold
with boundary. 
Then there is a geometrically natural choice of a parameter $T$
and, for $T$ sufficiently large, a one-parameter family of
solutions $(M_T, \Gt, K_{T})$ of the Einstein constraint equations
with the following properties. The $n$-manifold $M_T$ is
constructed from $\tilde M$ by adding a neck
connecting the two points $p_1$ and $p_2$. For large
values of $T$, the Cauchy data $(\Gt, K_{T})$ is a small
perturbation of the initial Cauchy data $(\tilde \gamma, \tilde
K)$  away from small balls about the points $p_a$.  In fact, for
any
$\epsilon>0$ 
and $k\in \N$ we have $(\Gt, K_{T})\rightarrow (\tilde \gamma,
\tilde K)$ as $T\rightarrow\infty$  in $C^{k}\left( \overline{M
\setminus \left(B(p_1,\epsilon)\cup
B(p_2,\epsilon)\right)}\right)$.
\footnote{One should note the absence of any nondegeneracy condition in
Theorem~\ref{Tggluing}.
As is evident in the proof, this is accounted for by the
imposition of Dirichlet boundary conditions.}
\end{Theorem}

\noindent{\sc Proof:} These solutions are constructed via the
conformal method following the technique developed in~\cite{IMP}.
The adaptation of the proof of Theorem~1 of~\cite{IMP} to allow
for initial data on manifolds with boundary requires only minor
variations which we indicate here. The construction begins with a
conformal deformation of the initial data within small balls about
the points $p_a$, $a=1,2$. The metric is conformally deformed to
make deleted neighborhoods of these points asymptotically
cylindrical. One then truncates these neighborhoods at a distance
$T$ (in the asymptotically cylindrical metric, for $T$ large) and
identifies the remaining ends to form the new manifold $M_T$ with
metric $\gt$. The first variation in the proof occurs when
deforming the approximate transverse--traceless $\mut$ formed by
gluing  the conformally transformed $\tilde \mu$ across the neck via
cut-off functions. This requires solving (with appropriate
estimates) the elliptic system
$$
LX=W
$$
where $W=\dive_{\gt}\mut$ is supported near the center of the 
asymptotic cylinder,
$L=-\dive_{\gt}\circ{\cal D}$ and ${\cal D}X= \frac12{\cal L}_X\gt
-\frac1n(\dive_{\gt} X)\gt$ is the conformal Killing operator
applied to the (unknown) vector field $X$. In~\cite{IMP} the required
uniform invertibility of $L$ is established under a nondegeneracy
condition which amounts to the absence of conformal Killing
vectors fields (which are in the kernel of $L$) vanishing at
$p_a$. When $\tilde M$ has a non-empty boundary we are actually
interested in solutions to the boundary value problem
\be\label{vlapb}
\left\{
\begin{array}{rll}
LX&=&W \quad \hbox{ in }\quad M_T\\
X&=& 0 \quad \hbox{ on }\quad \partial M_T.
\end{array}
\right. \ee 
The core to solvability of  this problem is provided
by Theorem~2 of~\cite{IMP}. The proof in the present setting is identical
to the one there with the exception that the step where the
nondegeneracy condition (Definition~1 of~\cite{IMP}) is evoked is
now replaced by the nonexistence of conformal Killing fields which
vanish on the boundary, $\partial \tilde M$ (see, {\em e.g.},\/
Proposition~6.2.2 of~\cite{AndChDiss}). The required estimates on
the solution follow from Corollary~1 of~\cite{IMP} coupled with
the boundary Schauder estimates.  Setting $\sigma_T ={\cal D}X$
and $\tilde{\mu}_T =  \mu_T- \sigma_T$, we see that
$\tilde{\mu}_T$ is our desired transverse-traceless tensor.

The other modification occurs in solving the nonlinear boundary
value problem
\be\label{lich-bvp}
\left\{
\begin{array}{rll}
{\cal N}_{T}(\pt +\eta_T)&=&0  \quad \hbox{ in }\quad M_T\\
\eta_T &=& 0 \quad \hbox{ on }\quad \partial M_T.
\end{array}
\right. \ee where $\eta_T$ is presumed to be a small perturbation of an
explicit approximate solution $\pt$, and ${\cal N}_{T}$ is the
Lichnerowicz operator.
\be
\begin{array}{rlll}
{\cal N}_T(\psi) &=& \displaystyle{
\Delta_T \psi - \frac{n-2}{4(n-1)} R_T \psi} &+
\displaystyle{\frac{n-2}{4(n-1)}
|\tilde{\mu}_T|^2 \psi^{\frac{-3n+2}{n-2}}}\\ 
&&& \displaystyle{- \frac{n-2}{4n}\tau^2 \psi^{\frac{n+2}{n-2}}.}
\end{array}
\label{licht} \ee
Equation (\ref{lich-bvp}) is solved by means
of a contraction mapping argument.  The key ingredient is a good
understanding of the linearised operator ${\cal L}_T$ on
$M_T$. ${\cal L}_T$ is the operator
\begin{equation}
\begin{array}{rlll}
{\cal L}_{T} &=& \displaystyle{\Delta_{\gamma_T} 
-\frac{n-2}{4(n-1)}\Big(R(\gamma_T)
}&\displaystyle{+\frac{3n-2}{n-2}\,|\tilde{\mu}_T|^{2}\,\pt^{-\frac{4(n-1)}{n-2}}}\\
&&& \displaystyle{+ \frac{(n-1)(n+2)}{n(n-2)} \,\tau^2\,\pt^{\frac{4}{n-2}}
\Big).} \label{linlich}
\end{array}
\end{equation}
The basic point is to show that, corresponding to the solutions to
the boundary value problem
$$
\left\{
\begin{array}{rll}
{\cal L}_{T}\eta&=&f  \quad \hbox{ in }\quad M_T\\
\eta &=& 0 \quad \hbox{ on }\quad \partial M_T,
\end{array}
\right. $$ 
we have an isomorphism between certain weighted
H\"older spaces on $M_T$ where the weight factor controls decay/growth
across the neck, and moreover for a certain range of weights,
there is a $T_0$ such that this map has a uniformly bounded inverse for
all $T\geq T_0$.  The proof of this follows \S5 of~\cite{IMP} and
relies on the fact that the boundary value problem
\be\label{linlich-soln-bvp}
\left\{
\begin{array}{rll}
\Delta_\gamma\eta- \left(|\mu|^2_{\gamma}
+\frac{1}{n}\tau^2\right)\eta &=&0
\quad \hbox{ in }\quad \tilde M\\
\eta &=& 0 \quad \hbox{ on }\quad \partial \tilde M
\end{array}
\right. \ee has no non-trivial solutions.  The linear operator
appearing in (\ref{linlich-soln-bvp}) is precisely the linearised
Lichnerowicz operator about the original solution $(\tilde
M,\tilde \gamma, \tilde K)$.

Letting $\tpt=\pt+\eta_T$ be the solution to (\ref{lich-bvp}) one finds
that the
desired solution to the constraint equations is then given by
$$
\Gt= \tpt^{\frac{4}{n-2}}\,\gt \qquad \mbox{ and }\qquad K_{T} =
\tpt^{-2}\,\tmut +\frac{1}{n}\tau\tpt^{\frac{4}{n-2}}\gt.
$$
The fact that these solutions converge uniformly to the original
initial data sets in $C^{k,\alpha}({\overline{\tilde M}})$ away
from small balls about the points $p_1, p_2$ follows from the
calculations of \S8 of~\cite{IMP} together with the boundary
Schauder estimates. \qed

The gluing construction of \cite{IMP2}, which
only requires the initial data to have constant mean curvature in 
small balls about the points
at which the gluing is to be done, also easily generalizes to manifolds with
boundary.  To show this, we need to introduce the notion of
{\it nondegeneracy} for solutions of the constraint equations which are not
necessarily CMC on manifolds with boundary. 
We do this in the context of the conformal method
for non CMC data, which works as follows:  Given a fixed background metric
$\gamma$, a trace-free symmetric tensor $\mu$, and a function $\tau$, 
if we can solve the coupled equations
\begin{eqnarray*}
\Delta_{\gamma}\phi - \frac{n-2}{4(n-1)}R_\gamma\phi
+ \frac{n-2}{4(n-1)}\big|\mu+\calD W\big|^2\phi^{\frac{-3n+2}{n-2}}-
\frac{n-2}{4n}\tau^2\phi^\frac{n+2}{n-2} &=& 0\\
LW -(\dive\mu-\frac{n-1}{n}\phi^{\frac{2n}{n-2}}\nabla\tau) &=& 0
\end{eqnarray*}
for a positive function $\phi$ and a vector field $W$, then the initial data
\begin{equation*}
\tilde{\gamma} =  \phi^{\frac{4}{n-2}}\gamma,
\qquad \widetilde{K} = \phi^{-2}(\mu+\calD W) +
\frac{\tau}{n}\phi^{\frac{4}{n-2}}\gamma,
\label{eq:newdata}
\end{equation*}
satisfies the ($\Lambda=0$) vacuum Einstein constraints (\ref{ce20n0vac})-(\ref{ce20n1vac}).
The first of these is again referred to as the Lichnerowicz equation.
We write this coupled system as $\calN(\phi,W;\tau) = 0$.
The mean curvature $\tau$ is emphasized here, while the
dependence of $\calN$ on $\gamma$ and $\mu$ is suppressed.
We are interested here in the boundary value problem
\be\label{IMP2-bvp}
\left\{
\begin{array}{rll}
\calN(\phi,W;\tau) &=&0  \quad \hbox{ in }\quad \tilde M\\
\phi &=& 1\quad \hbox{ on }\quad \partial \tilde M\\
W &=& 0\quad \hbox{ on }\quad \partial \tilde M.
\end{array}
\right. \ee

The linearization $\calL$ of $\calN$ in the directions $(\phi,W)$
(but not $\tau$) is of central concern.  We consider
this linearization relative to a specified choice of Banach spaces $\X$ and $\Y$,
each consisting of scalar functions and
vectors fields vanishing on the boundary.
If our manifold is not compact then one would also build into $\X$ and $\Y$
appropriate asymptotic conditions.

\begin{definition}
\label{nondg}
A solution to the constraint equation boundary value problem
(\ref{IMP2-bvp}),
is nondegenerate with respect to the Banach spaces $\X$ and $\Y$
provided $\calL: \X \to \Y$ is an isomorphism.
\end{definition}

The main result of the first gluing paper \cite{IMP} shows that
any two nondegenerate solutions of the vacuum constraint equations
with the same constant mean curvature $\tau$ can be glued.
For compact CMC solutions on manifolds without boundary,
nondegeneracy is equivalent to $K\not\equiv 0$
together with the absence of conformal Killing fields, while
asymptotically Euclidean or asymptotically
hyperbolic CMC solutions are always nondegenerate (cf. \S7 of \cite{IMP}).
In \cite{IMP2}, using a definition of nondegeneracy similar to that stated
above, we show how to glue non-CMC initial data sets, provided the data is CMC (same constant)
near the gluing points. The argument from \cite{IMP2} readily applies to similar sets of non-CMC
data on manifolds with boundary which are nondegenerate in the sense of Definition~\ref{nondg},
yielding the following;

\begin{Theorem}
\label{IMP2gluing} Let  $(\tilde M, \tilde \gamma, \tilde K; p_a)$
be a smooth, marked solution of the
Einstein vacuum
constraint equations with cosmological constant $\Lambda=0$
on $\tilde M$, an $n$-manifold
with boundary. We assume that the solution is nondegenerate and
that the mean curvature, $\tau=\tr_{\gamma}K$ is constant in the
union of small balls (of any radius) about the points $p_a$,
$a=1,2$. Then there is a geometrically natural choice of a
parameter $T$ and, for $T$ sufficiently large, a one-parameter
family of solutions $(M_T, \Gt, K_{T})$ of the Einstein constraint
equations with the following properties. The $n$-manifold $M_T$ is
constructed from $\tilde M$ by adding a neck connecting the two
points $p_1$ and $p_2$. For large values of $T$, the Cauchy data
$(\Gt, K_{T})$ is a small perturbation of the initial Cauchy data
$(\tilde \gamma, \tilde K)$  away from small balls about the
points $p_a$.  In fact, for any $\epsilon>0$ 
and $k\in \N$ we have $(\Gt, K_{T})\rightarrow (\tilde \gamma,
\tilde K)$ as $T\rightarrow\infty$  in $C^{k}\left( \overline{M
\setminus \left(B(p_1,\epsilon)\cup
B(p_2,\epsilon)\right)}\right)$.
\end{Theorem}

\section{The (global) gluing construction for the Einstein-matter 
constraints for manifolds with boundaries}
\label{Scsnv}

We need to show that the  gluing construction which we have just
described for the Einstein vacuum constraints on a manifold with
boundary can be extended to the case of the Einstein-matter
constraints (\ref{ce20n}), including a cosmological constant
$\Lambda$. In \cite{IMaxP}, Isenberg,
Maxwell and Pollack describe in detail how to carry out gluing
constructions analogous to that of \cite{IMP} for solutions of the
constraints for Einstein's theory coupled to a wide variety of
source fields (Maxwell, Yang-Mills, fluids, \emph{etc.}) on
complete manifolds. Here, we briefly describe how this works, and
we adapt these results to the case of a manifold with boundary.
For present purposes, we ignore any extra constraints which might
have to be satisfied by the matter fields, and we describe those
fields exclusively in terms of their stress-energy
contributions\footnote{This can be interpreted as a perfect fluid
model. However, we are not making any hypotheses upon the dynamics
of the matter fields.}  $\rho$ and $J^i$. These are required to
satisfy the dominant energy condition (\ref{dec}). We also allow
for the inclusion of a non-zero cosmological constant $\Lambda$.

So we start with a set of initial data $(\tilde M,\tilde \gamma,
\tilde K, \tilde \rho, \tilde J, \tilde \Lambda)$ which satisfies
the constraint equations (\ref{ce20n})-(\ref{ce20n1}) on $\tilde
M$, an $n$-dimensional manifold with smooth non-empty boundary. We
presume that this set of data has constant mean curvature $\tilde
\tau$, from which it follows that if we do a trace decomposition
$\tilde K=\tilde \nu + \frac{1}{3} \tilde \tau \tilde \gamma$,
with $\tilde \tau =\tr_{\tilde \gamma}\tilde K$, the trace-free
field $\tilde \nu$ satisfies the condition
$$
\tilde \nabla_j \tilde \nu ^{ji} = 8 \pi \tilde J^i.
\label{nu-eqn}
$$
Along with the initial data, we specify the pair of points $p_1,
p_2 \in M$ at which we carry out the gluing.

We recall that the first step of the gluing construction for
the vacuum constraints in \cite{IMP} involves a conformal blowup of the
gravitational fields at each of the points $p_1$ and $p_2$,
followed by gluing these fields together using cutoff functions
along the join of the two asymptotic cylinders created by this
blowup.
For the non vacuum case, we need to conformally transform
and glue the matter quantities $\rho$ and $J$ as well. The
conformal transformations which keep the conformal constraints
semi-decoupled for CMC data, which preserve the dominant energy
condition, and which lead to the simplest form for the
Lichnerowicz equation,  are $\tilde \rho \rightarrow \phi^{\frac{2n+2}{n-2}}
\tilde \rho$ and $\tilde J^i \rightarrow \phi^{\frac{2n+4}{n-2}} \tilde J^i$
(coupled with $\tilde \gamma_{ij} \rightarrow \phi^{-\frac{4}{n-2}} \tilde
\gamma_{ij},\ \tilde \nu_{ij} \rightarrow \phi^2 \tilde \nu_{ij}$,
and $\tilde \tau \rightarrow \tilde \tau$, together with $\tilde
\Lambda \rightarrow \tilde  \Lambda$.) As for the gluings, we
also apply a simple cutoff function procedure to 
$\phi^{\frac{2n+2}{n-2}} \tilde\rho$ and 
$ \phi^{\frac{2n+4}{n-2}} \tilde J^i$, thereby producing the smooth
fields $\tilde \rho_T$ and $\tilde J_T$ on $M_T$, along with
$\tilde \gamma_T, \tilde \nu_T$, and the constants $\tau =\tilde
\tau$ and $\Lambda=\tilde \Lambda$.

The next step is finding the traceless tensor $\tilde  \sigma_T$
which satisfies the momentum constraint
\begin{equation}
 \nabla^{({{\gamma_T}})}_j \tilde \sigma_T ^{ji} = 8 \pi
\tilde J_T^i. \label{sigma-eqn}
\end{equation}
Here $ \nabla^{({{\gamma_T}})}$ is the Levi-Civita covariant
derivative of the metric $\gamma_T$. We obtain $\tilde\sigma_T$ by
solving the boundary value problem
\be
\left\{
\begin{array}{rll}
LX&=&V \quad \hbox{ in }\quad M_T\\
X&=& 0 \quad \hbox{ on }\quad \partial M_T.
\end{array}
\right. \ee 
with $L$ as in Section~\ref{Sbound}, and with
$$
V=J_T - div_{ \gamma_T}\tilde \nu_T,
$$
and then setting $\tilde \sigma_T=\tilde \nu_T + {\cal D}X$
(recall that $\cal D$ has been defined in the paragraph preceding
\eq{vlapb}). Noting that $V$ is supported near the points $p_1$
and $p_2$, we readily verify that the arguments for solvability of
the boundary value problem (\ref{vlapb}) in Section 2 apply here as well.
We also obtain the required pointwise estimates for $\tilde
\sigma_T - \tilde \nu_T$.

The remaining step in the gluing construction of \cite{IMP} involves solving
the Lichnerowicz equation and then obtaining the requisite
estimates for the solution $\psi_T$ (ie, showing that away from
the neck, $\psi_T \rightarrow 1$ in a suitable sense).
For the Einstein-matter
constraints (\ref{ce20n})-(\ref{ce20n1}),
with the decompositions described here,
the Lichnerowicz operator takes the form (compare (\ref{licht}))
\bean\lefteqn{
{\cal N}_T(\psi) = \Delta_T \psi 
- \frac{n-2}{4(n-1)} R_T \psi +
\frac{n-2}{4(n-1)} |\tilde{\sigma}_T|^2 \psi^{\frac{-3n+2}{n-2}} }
&&
\\ &&\phantom{xx}
+ \frac{4 \pi(n-2)}{n-1} \rho_T\psi^{-\frac{n}{n-2}} 
- (n-2)\Big(\frac{1}{4n}\tau^2-\frac{1}{2(n-1)}\Lambda \Big)
\psi^{\frac{n+2}{n-2}}. \label{lichmat} \eea The matter-related
term in (\ref{lichmat}), $2 \pi \rho_T \psi^{-3}$, causes very few
changes in the analysis. We note, for example, that in the
expression for the linearised Lichnerowicz operator
\begin{eqnarray*}
\lefteqn{
{\cal L}_{T} = \Delta_{\gamma_T} -\Big(
\frac{n-2}{4(n-1)}R(\gamma_T)
+\frac{-3n+2}{4(n-1)}\,|\tilde{\sigma}_T|^{2}\,\pt^{\frac{4(-n+1)}{n-2}}}
&&
\\ &&\phantom{xxxxx} - \frac{4\pi n}{n-1} \rho_T \psi_T^{-\frac{2n-2}{n-2}}+
(n+2)\Big(\frac{1}{4n}\tau^2-\frac{1}{2(n-1)}\Lambda \Big)
\pt^{\frac{4}{n-2}} \Big),
\end{eqnarray*}
the $\rho$ term has  very much the same
effect as does the $\sigma$ term, so its presence does not alter
the proof of the existence of a solution  or the subsequent error
analysis.

The constant $\Lambda$, on the other hand, can cause trouble.
However the argument presented in Section 2 shows that ${\cal
L}_{T} $  has a uniformly bounded inverse for $T$ sufficiently
large provided that\bel{taulamc} (\tr_g K)^2 \ge \frac
 {2n}{(n-1)}
 \Lambda\;,\ee If this
condition holds, then the rest of the analysis goes through. We
thus have, finally

\begin{Theorem}
\label{Tggluingmat} Let  $(M, \gamma, K, \rho, J, \Lambda; p_1,
p_2)$ be a smooth, marked, constant mean curvature solution of the
Einstein matter  constraint equations on $M$, an $n$-manifold with
boundary. We assume that \eq{taulamc} holds, and that the dominant
energy condition $\rho \geq |J|$ holds. Then there is a
geometrically natural choice of a parameter $T$ and, for $T$
sufficiently large, a one-parameter family of solutions $(M_T,
\Gt, K_{T}, \rho_T, J_T, \Lambda)$ of the Einstein constraint
equations with the following properties. The $n$-manifold $M_T$ is
constructed from $M$ by adding a neck connecting the two points
$p_1$ and $p_2$. For large values of $T$, the Cauchy data $(\Gt,
K_{T}, \rho_T, J_T, \Lambda )$ is a small perturbation of the
initial Cauchy data $(\gamma, K, \rho, J, \Lambda)$  away from
small balls about the points $p_1, p_2$. In fact, for any
$\epsilon>0$ 
and $k\in \N$ we have $(\Gt, K_{T}, \rho_T, J_T, \Lambda
)\rightarrow (\gamma, K, \rho, J, \Lambda)$ as
$T\rightarrow\infty$  in $C^{k}\left({\overline M}\setminus
\left(B(p_1,\epsilon)\cup B(p_2,\epsilon)\right)\right)$.
\end{Theorem}

We note, without further discussion, that one can also readily produce a theorem analogous
to Theorem \ref{IMP2gluing}, but with matter included in the constraint equations.

\section{Proof of Theorems~\ref{Tlgluingv} and \ref{Tlgluingnv}}\label{Sgluing}

In the vacuum case let $(\mcM,g)$ be the maximal globally
hyperbolic vacuum development of the initial data $(\tilde
M,\tilde \gamma, \tilde K)$; in the non-vacuum case let $(\mcM,g)$
be the development of the data, the existence of which has been
assumed. In the vacuum case $\tilde M$ is achronal in $\mcM$ by
construction; in the non-vacuum case this can be achieved, without
loss of generality, by passing to a subset of $\mcM$.

 There exists $r_0>0$ such that for all $0<r\le r_0$,
 the open geodesic balls $B(p_a,r)$ in $(\tilde
M,\tilde \gamma)$ have smooth boundaries and   relatively compact
domains of dependence in $(\mcM,g)$. In the setting of
Theorem~\ref{Tlgluingnv} we set $\Omega_a=B(p_a,r_0)$. By reducing
$r_0$ if necessary we can assume that $\rho>|J|$ on the domains of
dependence $\mcD(\Omega_a)$. Without loss of generality we can
further assume that $r_0\le \epsilon/2$, where $\epsilon$ is the
radius chosen in the statement of the theorems.

By a result in\cite{ChBeignokids}, we can make an $\epsilon$-small deformation of
the initial data, supported in $\Omega_1\cup\Omega_2$, such that
the deformed initial data set satisfies the dominant energy condition,
remains vacuum if it was to begin with, still satisfies
$\mcK(B(p_a,r_0))=\{0\}$, and now moreover there exists an $R$
such that for every $r_-$ and $r_+$ satisfying $0<r_-<r_+<R<r_0$,
we have
\bel{nokidp}\mcK(\Gamma(p_a,r_-,r_+))=\{0\},
\ee where $\Gamma(p_a,r_-,r_+):= B(p_a,r_+)\setminus
\overline{B(p_a,r_-)}$. (In fact, the deformation can be arranged
so that $\mcK(\mcU)=\{0\}$ for any open set 
$\mcU\subset B(p_a,r_0)$.) In
vacuum,
replacing $\Omega_a$ with $B(p_a,r_0)$ if necessary, we may
work in  $B(p_a,r_0)$ with $r_0$ being taken as small as desired.
We assume in what follows that this is the case.

For any  set $\Omega$ with a distance function $d$, we define
$$\Omega(s):=\{p\in \Omega: d(p,\partial\Omega)< s\}\;; $$
the sets $\Omega$ considered here will always be equipped with a
Riemannian metric, and then
 $d$ will be taken to be the distance function associated with this metric. In particular
we thus have $\Omega_a(s)=\Gamma(p_a,r_0-s,r_0)$.

Let us  denote by $(\gamma_a ,K_a)$ the initial data induced on
$\Omega_a$. We next wish to reduce
 the problem to that in which
$(\Omega_a,\gamma_a ,K_a)$ have constant (sufficiently large) mean
curvature. We choose a constant $\tau$ so that \bel{lambdain}
\tau^2 - \frac{2n}{(n-1)}\Lambda \geq0\;.\ee As the domains of dependence
$\mcD(\Omega_a)$ are compact, we can use the work of Bartnik
~\cite[Theorem~4.1]{bartnik:variational} to conclude that there
exist  smooth spacelike hypersurfaces $\hOmega _a\subset \mcM$,
with boundaries  $\partial \Omega_a$,  on which the induced data
$(\gamma_a, K_a)$ satisfy
\bel{consttr} \tr_{\gamma_a } K_a = \tau\;.\ee

In the Einstein matter case, with $\rho> |J|$, we appeal to the results
in~\cite{ChBeignokids} to obtain a small perturbation of the data
induced on $\widehat \Omega_a$, preserving \eq{consttr}, such that
there are no KIDs on any open subset of the regions $\widehat
\Omega_a$. By continuity the dominant energy condition $\rho>
|J|$ will still hold provided the perturbation is small enough.

In the vacuum case, we claim that
 the domains
${\widehat{\Omega}}_a$  have no local KIDs  on every collar
neighborhood of their boundary. Indeed, suppose that this is not
the case. Then there exists a collar neighborhood, say
${\widehat{\Omega}}_1(s)\subset {\widehat{\Omega}}_1$, with a non
trivial set of KIDs there. Therefore there exists a non-trivial
Killing vector field $X$ on the domain of dependence
$\mcD({\widehat{\Omega}}_1(s))$. But the intersection
$$\mcD({\widehat{\Omega}}_1(s))\cap \Omega_1$$
contains some collar neighborhood $\Omega_1(s_1)$, and therefore
$X$ induces a KID there, contradicting \eq{nokidp}.

 For all $s_0>0$ the argument just given also guarantees the
existence of an $s_1$ satisfying $0<s_1<s_0$ such that
\bel{nokidshat} \mcK({\widehat{\Omega}}_a(s_0)\setminus\overline{
{\widehat{\Omega}}_a(s_1)})=\{0\}\;.\ee

 The process described so far reduces the problem to one with
 CMC initial data satisfying \eq{lambdain}-\eq{nokidshat}, on
 a compact manifold with boundary.
(As pointed out in the introduction, the hypothesis of existence
of the associated space-time, made in Theorem~\ref{Tlgluingnv}, is
not needed for such data.) Choose, now, a pair of  points $\hat
p_a\in{\widehat{\Omega}}_a\setminus
\partial {\widehat{\Omega}}_a$. Applying Theorem~\ref{Tggluing} in vacuum or Theorem~\ref{Tggluingmat}
with matter to $({\widehat{\Omega}}_a,\gamma_a,K_a)$ for any
sufficiently small $\epsilon$,  we  obtain a glued initial data set
$(\widehat M,\gamma(\epsilon),K(\epsilon))$, where ${\widehat{M}}$
is the manifold, with boundary
$\partial{\widehat{M}}=\partial\hOmega _1\cup
\partial {\widehat{\Omega}}_2$, which is the connected sum of $\hOmega_1$
and $\hOmega_2$ across a small neck around the points $\hat p_a$.
Let $s_0>0$ be any number such that $\hat p_a\not \in
{\widehat{\Omega}}_a(s_0)$.  On ${\widehat{\Omega}}_a(s_0)$ the
deformed data $(\gamma(\epsilon),K(\epsilon))$ arising from
Theorem~\ref{Tggluing} approach $(\gamma_a,K_a)$ in any $C^k$ norm
as $\epsilon$ goes to zero.  As a consequence of  \eq{nokidshat},  the construction
presented in Section~8.6 of~\cite{ChDelay} can be carried through
at fixed $\rho$ and $J^i$  and it gives, for all $\epsilon$ small
enough, a smooth deformation of $(\gamma(\epsilon),K(\epsilon))$
on ${\widehat{M}}$, which coincides with $(\gamma_a ,K_a)$ on
${\widehat{\Omega}}_a(s_1)$, and coincides with
$(\gamma(\epsilon),K(\epsilon))$ away from $\hOmega _a(s_0)$. The
deformation preserves the strict dominant energy condition
(reducing $\epsilon$ if necessary), or is vacuum if the original
data were.

Consider, finally, the manifold $M$ which is obtained by gluing together $\tilde
M\setminus \left(\Omega_1\cup\Omega_2\right)$ and ${\widehat{M}}$,
across $\partial {\widehat{M}}$. $M$ carries an obvious initial
data set $(\gamma,K)$, which is smooth except perhaps at
the
gluing boundary $\partial{\widehat{M}}$, at which  both $\gamma$ and
some components of $K$ are at least continuous. But in a
neighborhood of $\partial {\widehat{M}}$, bounded away from the neck region,  the data $(\gamma,K)$
coincide with those arising from a continuous, piecewise smooth
hypersurface in $\mcM$, which consists of a gluing of $\tilde M$
on one side of $\partial {\widehat{M}}$, with
${\widehat{\Omega}}_a$ on the other. If we smooth out that
hypersurface in $\mcM$ around $\partial {\widehat{\Omega}}_a$, then the
new data near $\partial \Omega_a$ arising from the smoothed-out
hypersurface,  provides a smoothing  of the initial data
constructed so far.
 \qed

\section{Applications}
\subsection{Vacuum space-times without CMC surfaces}\label{Sex}

In~\cite{bartnik:cosmological} Bartnik has constructed an
inextendible spatially compact space-time, satisfying the dominant
energy condition, which has no closed CMC hypersurfaces (see
also~\cite{BrillMGM,Witt:1986ng,IMP2}). Here, using a construction
analogous to that proposed by Eardley and Witt~\cite{EW}, we prove
a similar result (Corollary~\ref{Tex}) for  {\em vacuum}
spacetimes. The key step is proving the existence of vacuum
initial data on a connected copy of $T^3$ with itself, with the
property that the metric is symmetric under a reflection across
the middle of the connecting neck, while $K$ changes sign under
this reflection. The non-existence of closed CMC surfaces in the
maximal globally hyperbolic development of those initial data
follows then from the arguments presented
in~\cite{bartnik:cosmological}.

 Let $\hat \gamma$ be any
metric on $M=T^3$ which has no conformal Killing vectors (such
metrics exist, \emph{e.g.} by \cite{ChBeignokids}), let
$\hat\mu\not \equiv 0$ be any transverse traceless tensor on $M$
(such tensors exist, \emph{e.g.} by~\cite{BergerEbin}), and let
$\hat K=\hat\mu+\tau \hat\gamma$, for some constant $\tau\ne 0$.
It follows, \emph{e.g.} from~\cite{Jimconstraints}, that the
conformal method applies, leading to a vacuum initial data set
$(\tilde \gamma, \tilde K)$, with $\tilde \gamma$ being a
conformal deformation of $\hat \gamma$. Now, it is easily checked
that for CMC data $(\tilde \gamma, \tilde K)$ on a closed
manifold, a KID $(N,Y)$ must have $N=0$ and must have $Y$ a
Killing vector field of $\tilde \gamma$. Consequently, $Y$ is a
conformal Killing vector for $\hat \gamma$, so that $Y=0$ by our
hypothesis on $\hat \gamma$. It follows that $(\tilde \gamma,
\tilde K)$ does not have any nontrivial global KIDs; \emph{i.e.,}
$\mcK(M)=\{0\}$.

Let $(\mcM,g)$ be the maximal globally hyperbolic development of
the data. As in Section~\ref{Sgluing},  we can deform the initial
data hypersurface in $\mcM$ to create a small  neighborhood of a
point $p$ in which the trace of the new induced $\tilde K$
vanishes, while maintaining the  condition $\mcK(M)=\{0\}$.
We use the same symbols $(\tilde \gamma, \tilde K)$ to
denote the new data.

Now let $\tilde M$ consist of two copies of $M$, with initial data
$(\tilde \gamma, \tilde K)$ on the first copy, say $M_1$, and with data
$(\tilde \gamma, -\tilde K)$ on the second copy, say $M_2$. We let
$\Omega_a=M_a$,  and we let $p_a$ denote the points in $M_a$ corresponding to $p$.
Noting that the mean curvature vanishes in symmetric neighborhoods of
$p_1$ and $p_2$, we now apply the construction for
Theorem~\ref{Tlgluingv} presented in Section~\ref{Sgluing}.  To produce the
desired initial data set on $T^3\#T^3$, it is crucial to verify that
all the steps are done with the
correct symmetry around the middle of the connecting neck. In particular, we must check
 that the glued solution
obtained from Theorem~\ref{Tggluing}, when applied to such initial
data, leads to a solution of the constraints which has the desired
symmetry: this is achieved by using approximate solutions with the
correct symmetry in the construction used to prove
Theorem~\ref{Tggluing}. The end result follows from the
uniqueness
 (within the given conformal class)
of the solutions obtained there. We thus have verified
Corollary~\ref{Tex}.

\subsection{Bray's quasi-local inner mass}
In \cite{Bray:preparation2} Bray defines a notion of ``inner mass"
for a surface $\Sigma$ which is outer-minimising with respect to
area in an asymptotically Euclidean initial data set $(M,\gamma, K)$
satisfying the dominant energy condition \eq{dec} (see also
\cite{ChBray}). Given a surface $\Sigma\subset M$ which is
outer-minimising with respect to a fixed asymptotically flat end
of $(M,\gamma)$,
 define the region $I$ `inside' $\Sigma$ to be
the union of the components of $M\setminus \Sigma$ containing all
the ends of $M$ except for the chosen one. The inner mass
$m_{inner}(\Sigma)$ is then defined to be the supremum of
$\sqrt{A/16\pi}$ taken over all fill-ins of $\Sigma$ (or
replacements of $I\subset M$ with initial data sets (of arbitrary topology)
which satisfy \eq{dec} and extend smoothly to $M\setminus I$, 
with the data on $M\setminus I$  unchanged)  where $A$ is the
minimum area  of surfaces in the fill-in needed to enclose all the ends 
other than the chosen `exterior' end. Note the similarity of this
definition to Bartnik's notion of quasi-local mass
\cite{Bartnik89,BartnikTsingHua}.

It is by no means clear that extremal data which realise
$m_{inner}(\Sigma)$ exist.  However Bartnik has observed that the
construction described in this paper results in the following:
\begin{Theorem}
\label{quasi-local} Suppose that  $(M,\gamma, K)$ is an
asymptotically flat initial data set which realises the inner mass
$m_{inner}(\Sigma)$ for an outer minimising surface $\Sigma\subset
M$.  Thus there is a surface (not necessarily connected) 
$S\subset I$, the interior region of
$(M,\gamma, K)$ relative to $\Sigma$, such that $A={\mbox
Area}(S)$ satisfies  $m_{inner}(\Sigma)=\sqrt{A/16\pi}$. If there
is an open set $\Omega\subset I$ satisfying
$\partial\Omega=S\cup\Sigma$, then there is at least one
non-trivial KID on $\Omega$ ie. $\mcK(\Omega)\not=\{0\}$. In
particular, in the time-symmetric case, $K\equiv 0$, the resulting
vacuum space-time is static in the domain of dependence of
$\Omega$.
\end{Theorem}

The proof of Theorem~\ref{quasi-local} is an immediate consequence of the
fact that were there to be no KIDS on $\Omega$, $\mcK(\Omega)=\{0\}$,
we could apply
Theorem~\ref{Tlgluingv} and locally glue in an additional black hole whose
apparent horizon  would contribute an additional area to $A$. This
would contradict
the assumption  that the original data was extremal for the inner mass.

\bibliographystyle{amsplain}

\bibliography{
../references/newbiblio,%
../references/newbib,%
../references/reffile,%
../references/bibl,%
../references/Energy,%
../references/hip_bib,%
../references/netbiblio,../references/addon}

\end{document}